# Bioconvection and Bloom in Porous environment and Its Assisted Technologies: A Review

Samarendu Biswas[1], Sachidananda Mahato[1], and Jayabrata Dhar[1]*

[1]Department of Mechanical Engineering, National Institute of Technology Durgapur, India 713209

## Abstract

Bioconvection, a phenomenon arising from the collective motion of motile microorganisms, plays a crucial role in shaping microbial distributions, fluid dynamics, and bloom formation in aquatic environments. While extensive research has explored bioconvection in open waters, a significant proportion of microbial life exists within porous habitats, including soils, sediments, and subsurface environments, where the interplay between microbial motility and structural heterogeneity remains less understood. This review synthesizes recent advancements in the study of bioconvective dynamics within porous media, emphasizing the impacts of various external parameters on microbial transport, and spatial organization in structured environments. Additionally, we explore the various scenarios where bloom formation interacts with porous ecosystems. We further discuss the implications of microbial interactions with porous substrates for bioremediation, health management, and industrial applications. A dedicated section explores genetic engineering approaches aimed at modulating microbial motility and growth characteristics to enhance bioconvective behaviour and bloom regulation. By integrating experimental, numerical, and theoretical insights, this review highlights key knowledge gaps and outlines future research directions toward a deeper understanding of microbial transport and bioconvection in porous systems, with potential applications in environmental sustainability and biotechnology.

*Corresponding email: jdhar.me@nitdgp.ac.in

# 1. Introduction

Microorganisms are ubiquitous to both natural ecosystems and technological settings, largely due to their remarkable adaptability and ability to thrive in a variety of aquatic habitats, such as rivers, oceans, wetlands, and sub-surface rocks (Sar *et al.*, 2019; Shu and Huang, 2022). Microbial organisms, including bacteria, marine and freshwater algae, and protists – microbes that have contributed to the regulation of the Earth's average temperature at geological timescales (Falkowski, 1994; Meyers, 1997) – additionally play significant roles in various technological industries, including bioremediation, food production, medicine, fertilizers, and alternative fuels, to name a few (Sharma *et al.*, 2019, 2021), thereby developing a sustainable ecosystem to mitigate global warming threats. The success of a microbial population, measured by its capability to colonize diverse habitats, relies on numerous evolutionary strategies (Giovannoni and Stingl, 2005; Ghoul and Mitri, 2016). Among these strategies, few microorganism species have developed the ability to move, *i.e. motile species*, in response to certain external stimuli like gravity, light, or concentration gradients of nutrients, gases and/or harmful toxins, providing an additional arsenal to their endeavour for survival (Alexandre, Greer-Phillips and Zhulin, 2004; Clarke, 2017). Motile microorganisms can actively or passively traverse diverse regions of the aqueous biosphere throughout their life cycle, whether in search of nutrients, as a bet-hedging strategy, to evade predators, or to locate suitable environments for colony growth (Alexandre, Greer-Phillips and Zhulin, 2004; Ben-Jacob *et al.*, 2016; Keegstra, Carrara and Stocker, 2022). Certain motile microorganisms exhibit an intriguing feature during their growth cycle known as *bioconvection* (Hill and Pedley, 2005). Bioconvection refers to the spontaneous instability that arises in a shallow suspension of single-celled organisms, driven by the collective movement of swimming microorganisms—such as algae, bacteria, or protists—that exhibit an overall upward motion bias. The collective motion causes episodic and unstable stratification that is caused by accumulation of motile cells at the upper layers resulting in visible patterns of convection and induces ambient fluid flow (Bees, 2020). In particular cases, bioconvection is identified as a significant factor leading up to algal blooms (Smayda, 1997; Taylor, Hondzo and Voller, 2021) – a condition of excessive growth of a dominant species in an ecosystem which often causes harmful effects in the surrounding environment. While phytoplankton and cyanobacteria are the dominant entities to delineate blooms in their life cycle, often referred to as harmful algal blooms or HABs (Grattan, Holobaugh and Morris, 2016), bacteria have also shown to exhibit such blooms within the soil (Fuentes *et al.*, 2016).

*Corresponding email: jdhar.me@nitdgp.ac.in

There is an abundance of motility studies in the context of bioconvection and blooms explored in open waters (oceans and freshwaters) (Bees, 2020) albeit a large portion of organisms are, in fact, found to reside in four of the 'big five' habitats including soil, the upper oceanic sediments, the deep continental and oceanic sub-surface that possess a porous characteristics (Kallmeyer *et al.*, 2012; Flemming and Wuertz, 2019; Hoshino *et al.*, 2020). The emergent feedback driven by the active microbial interactions within porous media is fundamental to understanding their survival strategies, bioremediation potential, and applications in health management (Podola, Li and Melkonian, 2017; Cui *et al.*, 2019; Scheidweiler *et al.*, 2019; Soares *et al.*, 2023; Jin and Sengupta, 2024). This has driven lab-based experimental studies to uncover the complexities of bioactive flows through artificial porous environments (Brun-Cosme-Bruny *et al.*, 2019, 2020; Théry *et al.*, 2021). In addition, convection induced by the collective motion of microbial communities at microscopic scales and bloom formation at population scales within porous systems may further reveal species-level emergence, resilience and crosstalk, facilitating varied survival strategies and technological applications. Thus, it is crucial to examine the key advancements in our understanding of motility patterns that drive bioconvection and contribute to bloom formation, shaped by the dynamic interplay between microbial entities and the porous environment across diverse ecosystems. Integrating these insights into modern technologies could offer a significant advantage in designing sustainable systems. To the best of the authors' knowledge, no comprehensive review currently addresses bioconvection and bloom formation within porous media, nor explores how these phenomena have been investigated in different contexts to enhance medium transport. This review provides a comprehensive synthesis of recent advances in the theoretical and experimental study of bioconvective dynamics within porous media and discusses the bloom formation capability of microorganisms in a porous environment. In addition, it presents a concise discussion on the genetic engineering approach aimed at modifying the attributes and phenotypic traits of algal growth characteristics and motility. The review is structured into four main sections: Section 2 explores recent experimental advancements in bioconvection in open waters; Section 3 summarizes the various numerical and experimental investigations of bioconvection in porous media; Section 4 evaluates the bloom formation potential of microbial communities in porous (or structured) environments; and Section 5 discusses advancements in genetic engineering that offer new avenues for controlling algal motility and physiological traits. Finally, the concluding remarks provides a summary of key findings and identifies future research directions in the field.

*Corresponding email: jdhar.me@nitdgp.ac.in

## 2. Bioconvection in Open waters:

Bioconvection refers to the spontaneous formation of convective patterns in a fluid due to the collective movement of motile microorganisms that are denser than the surrounding medium. The underlying mechanism of bioconvection is reminiscent of Rayleigh–Taylor instability (RTI) or Rayleigh–Bénard convection in miscible fluid pairs wherein a denser fluid layer residing above a less dense layer ensues an overturning hydrodynamic instability. In bioconvection, accumulation of non-neutrally buoyant cells at the upper surface induces an unstable stratification leading to active RTI-like phenomena that sustains for prolonged duration. RTI in passive fluid pairs is rigorously investigated in context of high energy technologies and processes (supersonic aviation, inertial confinement fusion), astrophysics, geophysics (Boffetta and Mazzino, 2017; Banerjee, 2020) and solubility trapping and long-term sequestration rate of carbon dioxide in the sub-surface aquifers (Dhar *et al.*, 2022; Soboleva, 2023). Recent findings further suggests that bioconvection may be attributed to unstable swimming orientation due to cell accumulation and flow gradients and has a length scale two to three order larger than the individual cell sizes (Williams and Bees, 2011; Bees, 2020). Moreover, metabolic activity has been shown to play a key role in the development of coherent flows within microbial suspensions (Fragkopoulos *et al.*, 2025). A shift in metabolism from photosynthesis to anaerobic respiration can alter single-cell motility, leading to the emergence of coherent flows at the population scale. In addition, while in classical RTI, the base state of the system remains unchanged, the base state of bioconvection is modified by the emergent flow (Hill and Pedley, 2005; Bees, 2020). Bioconvection of certain algae (*Chlamydomonas spp.*), bacteria (*Pseudomonas fluorescens* and *Chromatium okenii*) and protozoa (*Paramecium tetraurelia*) are observed in laboratory experiments (Kitsunezaki, Komori and Harumoto, 2007; Williams and Bees, 2011; Sato, Sato and Toyoshima, 2018; Ardré, Dufour and Rainey, 2019; Di Nezio *et al.*, 2023) while many natural settings (Sommer *et al.*, 2017; Sepúlveda Steiner, Bouffard and Wüest, 2021; Di Nezio *et al.*, 2023) also witness pattern swimming like bioconvection. Steiner et al. (Sepúlveda Steiner, Bouffard and Wüest, 2021) employed a rigorous 48-hour field experiment to investigate the persistent formation of bioconvective mixed layer in Lake Cadagno across the diel cycle. Despite midsummer stratification the authors observed substantial diapycnal mixing near the bioconvective mixed layer, suggesting effective interplay between turbulent diffusion and convection. Recent findings by Maldonado et al. (Torres Maldonado *et al.*, 2024) reveal that for bacterial concentration $\geqslant 0.45\%$ of swimming *E. coli*, induced bioconvection is strong enough to impact sedimentation of passive particles having size similar to that of the bacteria. In fact, Carvajal et al. (Martinez Carvajal, Taidi and Jarrahi, 2024) exploited the bioconvection exhibited by motile phototactic bacteria to enhance biomixing in an otherwise unstirred photobioreactor. The population POM (phototactic oscillatory motion) at scales equivalent to the bioreactor's width persisted for seven days without a mechanical stirrer. Recent observations on inverted bioconvection, configuration where the fluid density exceeded the cell density, has been experimentally and numerically investigated (Sato, Sato and Toyoshima, 2018; Alloui *et al.*, 2024). Simultaneously, bioconvection can be actively modulated by photo-bioconvection Rayleigh number and the beam width (employing heterogeneous light field), as observed by Ramamonjy et al. (Ramamonjy, Brunet and Dervaux, 2023), that exhibit self-organization, intriguing symmetry-breaking instabilities and enhance device integration. Gallardo-Navarro et al. further observed bioconvection in suspension comprising multiple species where species-specific motility behaviour dictates locally dominated segregation to optimize access to limited resources (Gallardo-Navarro *et al.*, 2025). Evidently, bioconvection in open waters have received due attention (Figure 1) and remains comprehensively reviewed in the last two decades with detailed review up till 2020 (Hill and Pedley, 2005; Bees, 2020). The present review therefore focuses primarily on bioconvection and blooms of microorganisms across diverse porous environments.

*Corresponding email: jdhar.me@nitdgp.ac.in

**Fig. 1:** Bioconvection patterns observed in different studies where panel (a) and (c) are top views while panel (b) and (d) are front views and panel (e) is the 3D view of the induced pattern. **(a)** Induced bioconvective pattern in a Petri dish (depth: 0.4 cm; width: 5 cm; cell density: $10^6$ cells cm$^{-3}$). The dish is illuminated from below with white light, while the lower half is covered with a red filter (660 nm). Distinct bioconvection patterns emerge: in the top half, white light induces upward swimming via phototaxis aligned with gravitaxis, suppressing gyrotaxis and triggering broad overturning instabilities; in contrast, cells in the red-illuminated bottom half exhibit no phototactic response and form tightly focused gyrotactic plumes (adapted from (Williams and Bees, 2011)). **(b)** Bioconvection plumes in a suspension of *Chlamydomonas augustae* having cell concentration greater than $10^5$ cells/cm$^3$ (adapted from (Williams and Bees, 2011)). **(c)** A suspension comprising five motile species, namely *Bacillus infantis*, *Rossellomorea aquimaris*, *P. megaterium*, *B. cereus* (labeled in green), and *Exiguobacterium sp.* (labeled in red) emerge bioconvective pattern (image captured 80 min after mixing, adapted from (Gallardo-Navarro *et al.*, 2025)). **(d)** Time-lapse lateral microscope images showing the emergence of inverted bioconvection in the central region of the cuvette (adapted from (Sato, Sato and Toyoshima, 2018)). Conditions: cell density = $1.0 \times 10^7$ cells/mL; light intensity = 1.4 μmol photons m$^{-2}$ s$^{-1}$. **(e)** 3D visualization of plume formation shown in panel (d), generated using VCAT5 software (adapted from (Sato, Sato and Toyoshima, 2018)).

## 3. Bioconvection in Porous Environments:

*Corresponding email: jdhar.me@nitdgp.ac.in

A porous medium is a material containing a network of interconnected voids (pores) within a solid structure, allowing the passage of fluids through its interconnected pore spaces. A porous medium or environment is characterized by porosity, permeability and tortuosity under the mean field approximation (Sahimi, 2011; Bear, 2013). In Darcy scale description porosity is the void fraction in the material, permeability quantifies the ease with which a fluid can flow through a porous material, and tortuosity describes the complexity of the fluid paths relative to a straight-line path through the medium. Most of the studies concerning bioconvection in porous medium has employed Darcy scale approximations or its derivatives and extensions. Bioconvection in porous media occurs when microorganisms, such as motile bacteria, protozoa or algae, migrate within a porous structure, inducing convective flows due to their own movement and environmental responses. Induced bioconvection can significantly influence the distribution and transport of substances, impacting processes such as nutrient cycling, pollutant dispersion, and sedimentation. The topic of bioconvection in porous environments have received attention primarily via theoretical observations and lacks strong experimental support. It is reported that bioconvection is likely to occur in porous systems that have permeability in order of $10^{-7}$ m$^2$ (Pedley and Kessler, 1992; Bees, 2020). Small pore size hinders free downwelling of bioconvective plumes and the obstructs the balance between gravitaxis and gyrotaxis that sustains a bioconvection. Recent study experimentally demonstrated that introduction of the porous layer altered the stability dynamics of the suspension. Without the porous layer, phototactic behaviour could lead to instabilities and bioconvection. However, with the porous layer in place, a more stable and concentrated layer of microalgae was achieved (Prakash and Croze, 2021). Porous environment with low permeability, thus, remains efficient to accumulate cells in high numbers resulting in enhanced harvesting capacity (Kessler, 1982). Although soil-based porous systems do not comply to the ranges of permeability that allows bioconvection, gravel-based regions within and around oceans and lakes showcases the possibility of bioconvection within their pores (Figure 2). Moreover, it is intriguing to note that the species *Chlamydomonas reinhardtii*, found in soil habitat (Harris, 2001; Sasso *et al.*, 2018), is known to exhibit bioconvection (Kage *et al.*, 2013; Sato, Sato and Toyoshima, 2018). Additionally, technological devices such as photobioreactors can employ porous systems that can exploit bioconvection patterns to enhance energy-efficient biomixing. In the wild, there are a few studies investigating transport and interaction of microorganisms with porous systems. Interfacial water flows resulting from the interaction of bottom currents and seabed topography provide a rapid and efficient mechanism for transporting suspended phytoplankton into the subsurface layers of porous sandy sediments, as demonstrated by Huettel et al. (Huettel and

*Corresponding email: jdhar.me@nitdgp.ac.in

Rusch, 2000). Field testing and flume experiments are used to illustrate these flows. This experiment aimed to investigate whether interfacial water flows arising from the interaction between boundary flow and topography promote the transport and destruction of suspended phytoplankton in sands with permeabilities like those of shallow shelf sediments. In their conclusion, they propose that permeable shelf sands function as efficient filters for particulate organic matter (POM) that simultaneously act as biocatalysts to hasten the recycling of nutrients and mineralization of organic carbon. According to Krajnc et al. (Krajnc *et al.*, 2022) the formation of a pellicle (floating biofilms - a structured, spatially heterogeneous, and porous-like extracellular matrix) at the water-air interface is connected to the vertical distribution and viability of individual B. subtilis PS-216 cells throughout the water column. Using a combination of real-time interfacial rheology and time-lapse confocal laser scanning microscopy, mechanical parameters were correlated to morphological changes of the growing pellicle. The results imply that pellicle development is a complex response to a changing environment caused by the growth of the bacteria. This response causes the model system's viable habitat to decrease to the water-air interface and causes cell development, morphogenesis, and population redistribution. A building with a mechanical stress-supporting structure eventually gets close to the limit thickness because to dietary constraint. Thus, understanding motility and bioconvection is essential for optimizing bioreactors, managing groundwater contamination, developing knowhow of biofilm growth and its environmental interactions and designing effective environmental remediation systems.

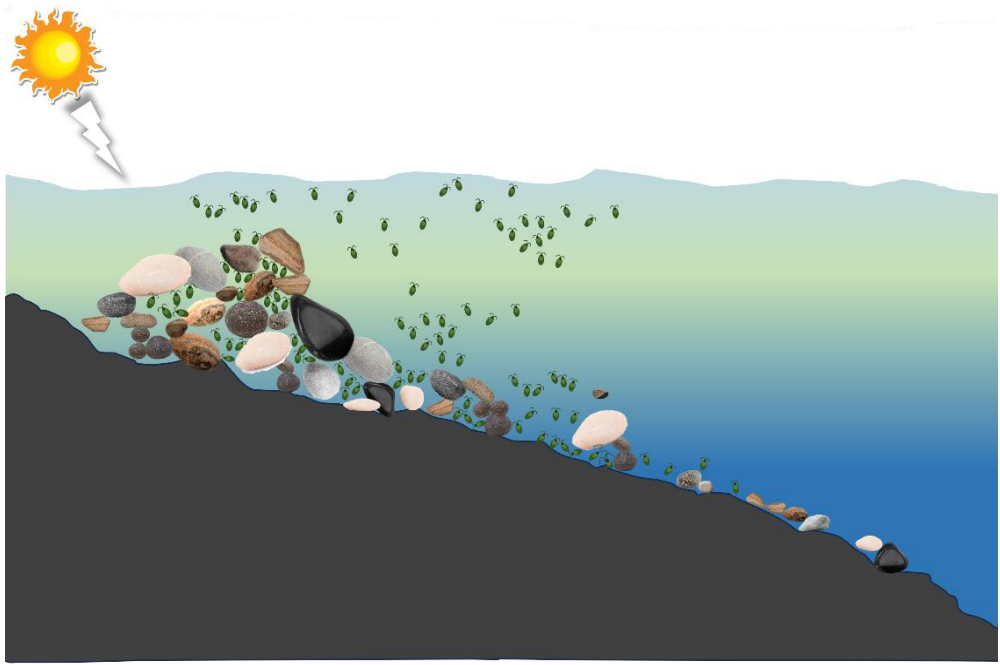

*Corresponding email: jdhar.me@nitdgp.ac.in

**Fig. 2:** A seabed or lakebed schematic where bioconvection and bloom can coexist among the porous regions around intertidal formation and rocky enclosures (created by authors).

Early theoretical studies of bioconvection in porous systems was carried out by Kuznetsov and co-workers (Kuznetsov and Jiang, 2001; Kuznetsov, Avramenko and Geng, 2004; Kuznetsov and Avramenko, 2005). His findings showed that bioconvection plumes with the highest downward filtration velocity of oxytactic bacteria effectively transport oxygen to the lower region of the chamber. The results also indicate that cell concentration increases toward the centre of the plume from its outer edges, whereas oxygen levels decrease in the same direction. Thereafter, numerous studies have attempted to theoretically and numerically address the influence of various factors such as bioconvection Rayleigh number ($Ra_b$), Lewis ($Le$), Péclet ($Pe$), Prandtl number ($Pr$) and Rayleigh numbers ($Ra$), to name a few, on the resulting bioconvection and its influenced parameters (Nusselt number, $Nu$, velocity characteristics, and Sherwood number, $Sh$) in diverse porous settings. All the symbols used and the dimensionless parameters that may affect the bioconvective dynamics is tabulated in Table 1 and 2, respectively. Majority of these studies employed the classical Darcy-scale formulation and its derivatives to simulate the instability. However, given bioconvection fundamentally differs from RTI, the present review addresses the various notable studies that investigates distinct factors influencing bioconvection to advance the field of microbial convection and motion patterns within porous environment.

## Table 1: Symbols

| | | | | | |
|---|---|---|---|---|---|
| $\boldsymbol{u}$ | Velocity field | $K$ | Permeability | $\gamma$ | mean volume of microorganism |
| $\rho_b$ | Density of nanofluid | $C$ | Oxygen concentration | $T_{ref}$ | Reference temperature |
| $\varepsilon$ | Porosity | $m$ | Microorganism concentration | $\phi$ | Nanoparticle concentration |
| $\sigma_b$ | Electrical conductivity | $k_b$ | Thermal conductivity of nanofluid | $\beta$ | Coefficient of thermal expansion |
| $B_0$ | Magnetic field strength | $T$ | Temperature field | | |
| $\mu_b$ | Dynamic viscosity of nanofluid | $D_c$ | Oxygen diffusivity | $(u, v)$ | Velocity components in the $(x, y)$ - directions |
| $W_c$ | Maximum cell swimming speed | $\delta$ | Oxygen utilization rate by cell number densoty | $m$ | Number density of motile microorganism |
| $D_m$ | Microorganisms' diffusivity | $\Delta \rho$ | $\rho_m - \rho_\omega$ | $\rho_m, \rho_\omega$ | Density of cell, base fluid |
| $C_{min}$ | Minimum oxygen concentration for cells to be alive | $T_\infty$ | Ambient temperature | b | Chemotaxis constant |

*Corresponding email: jdhar.me@nitdgp.ac.in

| g | Acceleration due to gravity (m/s$^2$) | L | Characteristic length | $D_n$ | diffusivity of microorganisms, $m^2 s^{-1}$ |
|---|---|---|---|---|---|
| $T_s$ | Surface temperature (K) | $\infty$ | Thermal Diffusivity $\frac{m^2}{s}$ | $K_f$ | Thermal diffusivity-based fluid |
| $W_l$ | Wall temperature | $k_c$ | Mass transfer coefficient | $v_{np}$ | Kinematic viscosity of the nanofluid |
| $k_c$ | Mass transfer coefficient | $\Delta T$ | Temperature different | $v_f$ | Kinematic viscosity of the fluid |
| D | Mass diffusivity | $\infty_{np}$ | Thermal diffusivity of the nanofluid | $\infty$ | Thermal diffusivity |
| $D_m$ | Microorganism diffusivity | $\frac{d\sigma}{dc}$ | Surface tension gradient concentration | $T_0$ | Reference temperature |
| $\Delta\rho$ | Density different | $\lambda$ | Fluid relaxation time | $\Phi$ | Viscous dissipation function |
| $\frac{d\sigma}{dT}$ | Surface tension gradient Temperature | S | Entropy generation | h | Convective heat transfer coefficient |
| $\gamma$ | Surface tension | R | Radius | $\rho_0$ | Reference Density |
| $\Delta P$ | Pressure difference | $D_c$ | Cell Diffusivity | $m_0$ | Average *m*. |

**Table 2: Dimensionless Numbers**

| Dimensionless Numbers | Expression |
|---|---|
| Prandtl number | $Pr = \dfrac{v_f}{\alpha_f}$ |
| Schmidt Number | $Sc = \dfrac{v_f}{D_c}$ |
| Grashof number | $Gr = \dfrac{g\beta_f(T_h - T_a)H^3}{v_f^2}$ |
| Hartmann number | $Ha = B_0 H \sqrt{\dfrac{\sigma_f}{\mu_f}}$ |
| Darcy number | $Da = \dfrac{K}{H^2}$ |
| Thermal Biot number | $Bi = \dfrac{Lh_j}{K_f}$ |
| Microorganism Biot number | $D_i = \dfrac{Lh_k}{D_n}$ |
| Concentration Biot number | $C_i = \dfrac{Lh_g}{D_b}$ |
| Richardson number | $Ri = \dfrac{Gr}{Re^2}$ |
| Lewis number | $Le = \dfrac{\alpha_f}{D_c}$ |
| Bioconvection Schmidt number ($Sc_b$) | $Sc_b = \dfrac{v_f}{D_m}$ |
| Richardson number (Ri) | $Ri = \dfrac{g(\Delta\rho)L}{\rho_0 U^2}$ or $= \dfrac{g\beta(\Delta T)}{U^2}$ |
| Reynolds number (Re) | $Re = \dfrac{\rho UL}{\mu}$ |

| Marangoni number (Ma) | $Ma = \dfrac{\dfrac{d\sigma}{dT} \times L \times \mu}{\infty \times \nu}$ |
|---|---|
| Weissenberg number (Wi) | $Wi = \dfrac{\lambda U}{L}$ |
| Bejan number (Be) | $Be = \dfrac{S_{heat}}{S_{heat} + S_{friction}}$ |
| Thermal biot number (Bi) | $Bi = \dfrac{h \times L}{k}$ |
| Curvature number | $K = \dfrac{\gamma L}{R \Delta P}$ |
| Bioconvection Lewis number ($Le_b$) | $Le_b = \dfrac{\infty}{D_m}$ |
| Bioconvection Biot number ($B_{ib}$) | $B_{ib} = \dfrac{h \times L}{D_m}$ |
| Sherwood number (Sh) | $Sh = \dfrac{K_c L}{D}$ |
| Nusselt Number (Nu) | $Nu = \dfrac{q_w \times L}{k_f(T_W - T_\infty)}$ |
| Bioconvection Péclet number ($Pe_b$) | $Pe_b = \dfrac{W_c \times L}{D}$ |
| Péclet number (Pe) | $Pe = \dfrac{UL}{D}$ |
| Nanoparticle Rayleigh number ($Ra_{np}$) | $Ra_{np} = \dfrac{g\beta_{np\Delta T}L^3}{\infty_{np}\nu_{np}}$ |
| Bioconvection Rayleigh number ($Ra_b$) | $\dfrac{g(\Delta\rho)L^3}{\nu D_m} = \dfrac{m_0 \Delta\rho\gamma}{\rho_f \beta_f \Delta T}$ |
| Rayleigh or Thermal Rayleigh number (Ra) | $\dfrac{g\beta(T_s - T_\infty)L^3}{\nu\infty}$ |

*Effect of thermal gradients*

Thermo-bioconvection has been a topic of key interest among many researchers who aims to influence the heat and mass transfer in a porous environment exploiting bioconvective dynamics of swimming microorganisms. Sharma et al. (Sharma and Kumar, 2011) theoretically studied the bioconvection in a dilute suspension of gyrotactic microorganisms in a horizontal shallow fluid layer saturated by a porous media and cooling from below, also a linear stability analysis is carried out. The results indicate that bioconvection only occur when the suspension becomes unstable due to an increase in temperature variance across the porous layer. The thermo-bioconvection in a square porous cavity filled with oxytactic microorganisms is studied numerically by Sheremet et al. (Sheremet and Pop, 2014). The momentum, heat and mass transfer in the porous region are solved using the Darcy model with the Boussinesq approximation, and the results show that the addition of oxytactic bacteria can result in a decrease of approximately 15% in Nusselt number (*Nu*) for low bioconvection Rayleigh number (*Ra*b) while a corresponding increase of approximately 9% is observed for high *Ra*b.

*Corresponding email: jdhar.me@nitdgp.ac.in

They further obtained distributions of velocity streamlines, isotherms, and the iso-concentrations of oxygen and motile microorganisms for different $Le$, $Pe$ and $Ra$. Balla et al. (Balla, Ramesh, *et al.*, 2020) investigates the effect of heat radiation in a square cavity filled with oxytactic bacteria. The results indicate that the intensity of bioconvection increases with $Ra$ and decreases with radiation parameter. The average $Nu$ and $Sh$ of microorganisms and oxygen density are enhanced by the $Ra$ and $Ra_b$. The study was extended to include the effects of chemical reaction by Balla et al. (Balla, Alluguvelli, *et al.*, 2020). They reported that $Ra$ determines the flow strength and increases the iso-concentrations of microorganisms and oxygen. A chemical reaction that enhances cell flow triggers cell division while weakening microorganism and oxygen iso-concentrations besides altering temperature distribution patterns. Naganthran et al. (Naganthran *et al.*, 2021) addressed the fluid flow and heat transfer problems related to free convection in micropolar (fluid with rotating microstructures) boundary layers using a vertical flat stretching plate in a porous medium. It is observed that as the micropolar parameter rises velocity and temperature reduces but microrotation profiles are enhanced, while higher Darcy numbers lead to increased velocity and temperature distributions. Belabid et al. (Belabid and Allali, 2021) investigated the thermo-bioconvection effects of phototactic microorganisms in a horizontal porous annulus filled with a Newtonian fluid and featuring wavy walls. The focus of the study is to understand the impact of the wall waviness on thermo-bioconvection and the distribution of motile bacteria within the porous annulus. Chowdhury and Rao (Chowdhury and Rao, 2023) studied thermo-bioconvective flow in an oxytactic microorganisms-filled square tilted porous cavity being sinusoidally heated. The findings focussed on the effective of cavity tiltedness on the resulting $Sh$ and $Nu$. Observations indicate that for higher $Ra$, the cavity's tilt increases the flow strength while for a larger $Ra_b$ at $Pe = 0.1$, the flow strength is weakened by the cavity's tilt.

*Effect of thermal gradients in nanofluids*

Many studies have incorporated multiple factors that impacts the overall bioconvective dynamics such as incorporation of nano- or micro-particles, and regions of external or internal heating into their system. There are experimental evidences of bioconvection-driven system enhancing biomixing in a particulate suspension (Kurtuldu *et al.*, 2011; Kage *et al.*, 2013; Wilhelmus and Dabiri, 2014; Dervaux, Capellazzi Resta and Brunet, 2017; Sato, Sato and Toyoshima, 2018; Martinez Carvajal, Taidi and Jarrahi, 2024). Many of these experimental studies are performed where the bioconvective medium (algal medium with particulate suspension) resides in a Hele-Shaw-like chamber (Tuval *et al.*, 2005; Bees and Croze, 2014; Hope *et al.*, 2016), which is often considered as a form of porous medium with a porosity of 1 and permeability of $h^2/12$ ($h$ being the chamber gap). In contrast, most studies concerning Darcy-scale porous medium are numerical in nature as reviewed hereunder. Sheremet et al. (Sheremet, Pop and Rahman, 2015) investigates natural convection heat transfer within 3D porous enclosure saturated with a nanofluid utilizing Buongiorno's model and accounts for Brownian diffusion and thermophoresis slip for nanoparticles. They observed that low *Ra* and *Le* combined with a high thermophoresis parameter led to significant non-homogeneous nanoparticle distributions within the cavity. Uddin et al. (Uddin *et al.*, 2016) analyses the impacts of Stefan blowing, thermal slip, second-order velocity slip, and microorganism species slip on the bioconvection boundary layer flow of a nanofluid over a horizontal plate surrounded by a porous medium, considering passively controlled boundary conditions. The results indicate that the velocity field is more sensitive to second-order slip than first-order hydrodynamic slip. While mass slip somewhat lowers the concentration of nanoparticles, it increases the microorganism density, temperature, and velocity. In addition, a positive *Gr* increases temperatures and nanoparticle concentrations, sped up the bioconvection nanofluid boundary layer flow but decreases the motile microbe density function. Saini and Sharma (Saini and Sharma, 2018) investigate how vertical throughflow affects the onset of nanofluid thermos-bioconvection in a porous media with gravitactic microorganisms. Findings reveal porosity delays the instability onset while the $Ra_{np}$, $Pe_b$, and $Le_b$ destabilize it. Balla et al. (Balla *et al.*, 2019) studied the impact of thermophoresis force and Brownian motion on the bioconvection heat transfer in a square porous cavity saturated with nanofluids and oxytactic microorganisms. The findings indicate that there is a significant correlation between the thermophoretic force, *Ra*, $Ra_b$, and $Pe_b$, the flow strength and temperature distribution. The *Le* and Brownian motion both dampens the flow strength while the boundary layer increases while the iso-concentration profiles are reduced by thermophoretic force. Nima et al. (Nima *et al.*, 2020) analyses free-forced bioconvective boundary layer flow adjacent to a vertical plate embedded in a Darcian porous media containing gyrotactic microorganisms towards modelling a PEM fuel-cell transport. The collective motion improved mixing in system contributing to more efficient thermal and solutal transport. Additionally, forced convection improved fuel cell efficiency for certain buoyancy ratio parameter, $Ra_b$, *Le*, $Pe_b$, and $Le_b$. Ahmad et al. (Ahmad, Ashraf and Ali, 2020b) investigates the mass and heat transport, including gyrotactic microorganisms and nanofluids via a porous medium over a nonlinear stretching sheet considering heat generation and chemical reactions. While a porous medium virtually has no effect on the number of motile bacteria, it dramatically increases shear stress. The high $Pe$ estimates are linked to microorganisms' low diffusivity, which reduces the motile bacteria's density profile. Later they (Ahmad, Ashraf and Ali, 2020a) examined the effect of heat radiation and incorporated a nonlinear contracting-stretching surface. They delineated that the skin friction and density of motile microorganisms in an environment with thermal radiation are slightly affected by the presence of porous media. The impact of heat sinks and sources on bioconvection of a nanofluid in a porous square cavity containing gyrotactic microorganisms is investigated by Jamuna et al. (Jamuna and Balla, 2021). The results show that $Pe$ enhances the concentration and bioconvection of microorganisms and nanoparticles. The heat generation/absorption parameter increases the average $Nu$ while decreasing the average $Sh$ of microbes and nanoparticles. The heat and mass transfer rates for the unsteady bioconvection in an iso-thermal vertical permeable cone immersed in a porous medium with chemical reaction is investigated by Rao et al. (Rao $et\ al.$, 2021). The findings indicate that the rescaled microorganism density decreases with increasing $Le_b$, rate of response, $Pe_b$, and $Bi_b$ while it increases with increasing buoyancy ratio and $Ra_b$. Thermophoresis and Brownian motion strongly influence the $Sh$ and microorganism density. Hussain et al. (Hussain, Aly and Alsedias, 2022) studied the flow of a suspension containing NEPCMs and the bioconvection of oxytactic bacteria in an omega-shaped porous enclosure. Results show that the iso-concentration of microorganisms and oxygen increases with a rise in the $Ra_b$. Additionally, the fusion temperature controls the position and intensity of the melting zone in the omega-shaped enclosure. Hussain et al. (Hussain, Raizah and Aly, 2022) extended the literature by including the impact of heat radiation and wavy surfaces on the bioconvection of oxytactic microbes and NEPCMs in a vertically wavy porous cavity. The outcome demonstrates that heat capacity, iso-concentration of microorganism and oxygen, and isotherms are adjusted in response to variations in thermal radiation parameters. A wavy wall's undulation number regulates the bioconvection flow, heat capacity, and streamlines inside the wavy cavity.

*Corresponding email: jdhar.me@nitdgp.ac.in

*Corresponding email: jdhar.me@nitdgp.ac.in

Under the impact of viscous dissipation, Ahmad et al. (Ahmad $et\ al.$, 2022) investigate the heat-mass transfer of solid nanoparticles and motile gyrotactic microorganisms under bioconvection setting through a permeable media (Fig 3). The outcomes show that an increase in the bioconvection Schmidt number ($Sc_b$) and bacterial motility increases the rate of microorganism diffusion. Porous medium is seen to affect the skin friction thereby viscous dissipation and density of microorganisms. Balla et al. (Balla $et\ al.$, 2022) investigate the streamlines, isotherms, and iso-concentrations of the nanoparticle and microorganisms for various inclination angle in thermo-bioconvection setting within a porous, square-shaped cavity. Their data indicates that the fluid flow intensifies with inclination until it reaches an angle of 30° degrees inclination, thereafter it begins to decrease. When the cavity is tilted more than 180° degrees (chamber turns counterclockwise), a reverse flow is observed. The microorganisms and nanoparticle concentration grow with the inclination. Later, Bodduna et al. (Bodduna, Balla and Mallesh, 2023) extended the effects of inclination on thermo-bioconvection in a porous square cavity containing nanofluid and gyrotactic microorganisms. They reported that $Ra_b$ increases the average values of $Sh$ for microbes, nanoparticles, and $Nu$. The stability of thermo-bioconvection caused by gravitactic bacteria within an anisotropic porous layer flooded with Jeffrey liquid was investigated by Garg et al. (Garg, Sharma and Jain, 2023b). Results suggests the flow initiation is destabilized by rising $Pe_b$ and concentrations of gravitactic microorganisms. The system has non-oscillatory instability due to the Jeffrey factor and mechanical anisotropy, both of which accelerate the formation of bioconvection patterns. Later, Garg et al. (Garg, Sharma and Jain, 2023a) observed a triply diffusive thermo-bio-convection instability trend when gyrotactic microorganism species are subjected to irregular internal heating. Results reveal the suspension acquires destabilizing features in both the oscillatory and stationary regimes due to the irregular internal heating. Gyrotactic microorganisms exhibit an increased generation of triply-diffusive thermo-bio-convective patterns. Because microorganisms swim quickly, heat transmission is enhanced and the bioconvection process is accelerated. Habibishandiz et al. (Habibishandiz, Saghir and Zahmatkesh, 2023) investigated the natural convection of oxytactic microorganisms saturating a square porous cavity with nanofluid. Various boundary conditions were applied to the sidewalls with an objective to report the heat transfer intensifier configuration with microorganisms. While constant temperature reduces average $Nu$ with increasing $Pe$ and $Ra_b$, periodic temperature distribution enhances the rate by 38%. The microorganisms' mass transfer increases for both boundary conditions when the $Ra_b$ and $Pe$ rise. Shah et al. (Shah *et al.*, 2023) investigates the dispersion of gyrotactic microbes and nanoparticles of Erying-Powell fluid flow in the presence of stratification, variable fluid properties and unequal bulk diffusion coefficients. The study addresses the dependence of the heterogeneous catalyst concentration distribution at the surface on the space-based internal heat generation parameter. Findings conclude that kinetic energy increases with growing values of the temperature-dependent thermal conductivity parameter, leading to a considerable increase in fluid temperature. Ismail et al. (Ismail and Bhadauria, 2023) examine the effects of internal heating in a Rivlin-Ericksen nanofluid under three distinct gravity modulation types in a horizontal layer of porous material.

*Corresponding email: jdhar.me@nitdgp.ac.in

*Corresponding email: jdhar.me@nitdgp.ac.in

The results show that internal heating destabilizes the system. In contrast, the system gets stabilized with an increase in Brinkman Dracy number, thereby, damping the mass and heat transfer. Kopp et al. (Kopp and Yanovsky, 2024) examines the behaviour of weakly nonlinear thermo-bioconvection under gravitational modulation and internal heating in a rotating porous medium filled with a Newtonian fluid. The results of linear bio-thermal convective instability indicate that the system becomes more stable as the $Da$ and Taylor numbers rise. A rise in the $Pe$, internal heating parameter, and modified bioconvective Rayleigh-Darcy number causes the system to become unstable.

**Fig 3.** Schematics of bioconvective flow configuration in a porous medium with thermal, concentration, and momentum boundary layer (adapted from (Ahmad *et al.*, 2022)) over a stretching sheet.

*Effect of thermal gradient and magnetic field in nanofluids*

Majority of studies that investigated the thermo-bioconvection of nanofluid and bioconvective microbes have included the impact of magnetic field via the magnetohydrodynamic (MHD) phenomena. Since there are too many studies on this topic, we attempt to highlight some of the well cited recent articles.

Balla and Naikoti (Balla and Naikoti, 2019) investigate how the MHD bioconvection heat transfer in a square cavity containing oxytactic bacteria is affected by a magnetic field. The cavity's flow strength is decreased by the $Le$ and magnetic field. The $Ra$, $Le$, and $Ra_b$ all improve the motile iso-concentration pattern. The average $Nu$ decreases with the magnetic field and increases with the $Pe$ and $Le$. Hussain et al. (Hussain and Geridonmez, 2022) extended the study to solve the mixed bioconvection in a trapezoidal porous cavity and subjected to inclined periodic magnetic fields. The findings demonstrate that a rise in $Ha$ weakens both mass and heat transfer. Their minimum values occur at an angle of $\theta = 0°$ at the unit period of periodic magnetic field, where the average $Sh$ and $Nu$ decrease maximally at $Ha = 100$. Biswas et al. (Biswas *et al.*, 2022) thereafter studied the combined thermo-bioconvection in a W-shaped porous cavity and found exhibits a better heat and oxygen mass transfer as compared to square and trapezoidal cavities. The findings further indicate that, because of shear stress, a clockwise circulation forms in the mixed (both forced and natural convection) thermo-bioconvection close to the upper sliding cold wall. The effects of the magnetic field and porous media on the motion of gyrotactic microorganisms in nanofluids are investigated by Kontukar et al. (Kotnurkar and Giddaiah, 2019). Gyrotactic bacteria were used to enhance the heat transfer and improve the nanofluids' stability and enhance mass transfer. The simultaneous flow and heat transfer of two nanoliquids (Williamson and Casson) over a porous substrate is investigated by Zuhra et al. (Zuhra *et al.*, 2020) accounting for buoyancy forces in conjunction with gyrotactic microorganisms, cubic autocatalysis chemical reaction and magnetic field presence. The findings show that the homogeneous reaction (far field) increases the density of motile microbes, but the heterogeneous coefficient (on surface), magnetic field parameter, porosity parameter, inertial parameter, and *Pe* inhibit its concentration. Velocity augments with *Wi* but dampens with magnetic field parameter, and *Ra*. The MHD mixed bioconvection in a Cu-water nanofluid is studied by Mandal et al. (Mandal *et al.*, 2021) in a top-wall-translating container with a wavy sidewall that is evenly heated. They observed that up to a certain number of ideal surface undulations, the curved surface improves mass and heat exchanges. $Ra_b$ has a direct impact on the developed secondary circulation, unlike *Ha*, *Da*, *Le*, *Pe*, and *Gr*. Using a typical, differentially heated, two-sided lid-driven cavity, Praveen et al. (Parveen *et al.*, 2021) investigates the peristaltic flow of a conductive nanofluid containing chemotactic microorganisms through an asymmetric channel examining the effects of wall slip, convective boundary conditions, magnetic fields, Joule heating, viscous dissipation, thermal radiation, Brownian motion, and thermophoresis on heat and mass transfer (Figure 4). The article reported that bioconvection caused by the microorganisms contributes to better fluid mixing and influences the thermal and concentration boundary layers. In addition, Joule heating and viscous dissipation lead to a rise in fluid temperature, while increasing the slip parameter reduces the velocity near the wall. Biswas et al. (Biswas, Datta, *et al.*, 2021) investigated MHD thermo-bioconvection within a differentially heated, two-sided lid-driven porous cavity, simultaneously accounting for the combined effects of thermal gradients, and magnetic fields. The findings demonstrate that the concentration of the motile bacteria is strongly influenced by the translating walls' direction and speed. The upper region of the cavity contains a higher concentration of oxygen and motile bacteria. Biswas et al. (Biswas, Manna, *et al.*, 2021) examine the MHD thermo-bioconvection in a porous cavity that is saturated with nanofluid. The outcomes demonstrate that the bottom wall's curvature has a significant impact on the flow, temperature distribution, oxygen and microbe iso-concentrations. The average *Nu* falls with the *Ha* and the $Ra_b$ but rises with the *Da* and the peak height of the curved bottom wall. The average *Sh* rises with the *Da* but falls with the curved wall's peak height and the *Ha*. Waqas et al. (Waqas *et al.*, 2021) explores the MHD bio-convection of Carreau-Yasuda nanofluid flow in the presence of heat radiation. The key findings include decreased velocity fields with increasing $Ra_b$, magnetic parameter and buoyancy ratio while a decrease in the temperature field is observed with increasing *Pr*. Later Irfan et al. (Irfan *et al.*, 2024) included the effects of Joule heating and chemical reaction to the above study. Later, thermos-bioconvection of Casson-Carreau nanofluid under nonlinear radiation and chemical reaction was studied (Saleem and Hussain, 2024).

*Corresponding email: jdhar.me@nitdgp.ac.in

*Corresponding email: jdhar.me@nitdgp.ac.in

**Fig 4.** Schematics of the problem addressing 2D bioconvection of a nanofluid that is peristaltically induced along flexible boundaries of a channel having a thermal and magnetic field, $H_0$ (adapted from (Parveen *et al.*, 2021)). In the above configuration $a_1$ and $a_2$ denotes half channel width from central axis while a and b denote the wave amplitude in upper and lower half channel, respectively and λ its wavelength.

Shamshuddin et al. (Shamshuddin *et al.*, 2022) aim to explore the MHD bioconvection flow of nanofluids on a stretched surface that is inhabited by microorganisms through numerical analysis using the Keller-Box approach. The findings demonstrate that the slip velocity, thermal buoyancy, and porosity increase the microorganism-nanofluid flow velocity, but induced electromagnetic force causes the flow velocity to decrease with increasing magnetic term. Except for the *Pr*, all other thermofluidic parameters promote heat propagation and diffusion. Hussain et al. (Hussain, Aly and Öztop, 2022) study the MHD bioconvection flow in a porous lid-driven cavity containing oxytactic bacteria, streamlined obstacles, and an Ag-MgO/water hybrid nanofluid. The results show that *Le* augments *Sh* but has no effect on heat transmission. The biggest heat transfer and lowest mass transfer are created for the highest *Ri*. Simplified obstacle dimensions can be employed as a control parameter and have an contrasting effect on mass and heat transfer. In this study, Habibishandiz et al. (Habibishandiz and Saghir, 2022) investigated the effect of microbes on MHD mixed convection of nanofluid within a vertical annular porous cylinder. The results show that the presence of microbes reduces the rate of heat transfer while in mixed convection, heat transport is increased by raising the *Ha*. Increasing the *Re*, *Pr*, *Le*, aspect ratio, and radius ratios; conversely, the *Ra*$_b$, *Pe*, and *Le* reduce the average *Nu*. *Ha* and aspect ratio improve average *Sh*, whereas *Ra*, *Pe*, *Le*, and radius ratio have the opposite effect on bioconvection. Study conducted by Alsedais et al. (Alsedais, Al-Hanaya and Aly, 2023) investigated the impacts of magnetic field on bioconvection in a porous annulus filled with oxytactic microorganisms and NEPCMs (nano-encapsulated phase change materials). They highlighted that the velocity field enhances with *Ra*$_b$ while it reduces with multiple embedded cylinders in the annular region. Thus, the number of cylinders inserted contributes to controlling the nanofluid flow in a cavity and facilitating cooling processes. To model the bioconvection of NEPCM in a porous cavity having hexagonal

*Corresponding email: jdhar.me@nitdgp.ac.in

configuration with a rotating four-pointed star, Aly et al. (Aly and Hyder, 2023) employed the incompressible smoothed particle hydrodynamics (ISPH) methodology towards a novel way of addressing fractional-time derivatives. The results show that enlarging the inner four-pointed star, $Ra$ and $Ra_b$ may significantly enhance bioconvection flow and nanofluid velocity, which can be applied in thermal industries and electronic devices. In addition, the addition of nanoparticles and an increase in Hartmann number cause the nanofluid's velocity to decrease. Rao et al. (Rao and Deka, 2023) examined the MHD thermos-bioconvection of hybrid nanofluid over stretched porous sheets under the impact of chemical processes and thermal radiation. The findings demonstrate that the magnetic field has a beneficial impact on the temperature, nanoparticle concentration, and microbial concentration profiles but negative effects on the velocity profile. As the porosity parameter is increased, the velocity profile shows a decreasing trend, while the temperature and microbe concentration profiles show an upward trend. The heat transmission and MHD bioconvective effects for nanofluid over a porous and extending wedge is numerically studied by Ali et al. (Ali *et al.*, 2023). As the wedge angle parameter increases, the thermal and momentum boundary layer thickness decreases. Increasing the wedge movement parameter increases the fluid velocity, but the fluid temperature behaves negatively. The $Nu$ and local skin friction show opposing trends as the wedge movement parameter rises. When porous media, nanofluids, and microorganisms are present, Hussain et al. (Hussain, Ertam, *et al.*, 2023) used an L-shaped adiabatic baffle in a square porous area with a partial heater to examine the regulation of heat and mass transfer as well as fluid flow. Increasing magnetic field dampens mass transfer but augments heat transfer and motile microorganisms' density. The higher the porosity and $Da$, the greater the heat transmission while mass transfer exhibits the reverse behaviour. The motile microorganism density is insensitive to lower $Da$ values. The Cu-water nanofluid's thermo-magneto-bioconvection enhances heat and mass transfer in wavy porous enclosure as reported by Biswas et al. (Biswas *et al.*, 2023) in context of cleaner energy applications. The result indicates that a double circulation due to the bioconvective parameter, $Ra_b$, gets generates near the heated wall. This new circulation's shape is dictated by the wall waviness parameter. On the other side, the higher $Pe$ and $Le$ diminish the secondary vortex while strengthening the primary vortex. Hussain et al. (Hussain, Öztop, *et al.*, 2023) examined the impacts of changing the lid motion on a mixed bioconvection of hybrid nanofluid in a semi-circular (one-sided) porous cavity in presence of uniform magnetic field. The findings indicate that for a given $Da$, a greater average $Nu$ is achieved for upward moving lid as compared to a downward moving one. Additionally, heat transport is reduced for all values of $Da$ as porosity increases. The location of the lid-

*Corresponding email: jdhar.me@nitdgp.ac.in

driven wall is correlated with the amount and intensity of circulating cells. Abbas et al. (Abbas and Khan, 2023) analyse the Marangoni convection effects on the MHD bioconvection of a Cross nanofluid over a sheet embedded in a porous medium, considering variable thermal conductivity, heat generation, and thermophoretic particle deposition. The study concluded that with Marangoni ratio parameter, the velocity profile improves, but temperature behaves in a contrasting manner. The skin friction coefficient decreases, and nanoparticle volume friction increases as the Marangoni number, $Wi$, and Marangoni ratio parameter increase. Ahmed et al. (Ahmed *et al.*, 2023) investigate the effects of binary chemical reactions, thermophoretic velocity, Soret-Dufour, and Joule heating phenomena on the unstable bioconvected Maxwell fluid flow. The results show that the velocity field, temperature, concentration, and dispersion of microbes all increased with an increase in the curvature parameter while the magnetic parameter decreased the velocity profile but increased the temperature, concentration, and microorganism profile due to the Lorentz force. Alsayegh (Alsayegh, 2024) extended to include the effect of higher-order slip in a triple diffusion flow of a viscoelastic nanofluid in a porous media. They found that the transport rates at the boundary decrease as the slip parameter and viscoelastic parameter rise. The application of the magnetic field increases skin friction and improves boundary layer cohesiveness due to its damping impact. The oscillatory bioconvective flow of a viscoelastic nanofluid was examined by Ghachem et al. (Ghachem *et al.*, 2024), considering the radiative effects, stretched, porous, oscillating surface and thermophoresis. The study reports that the maximal viscoelastic fluid constant increase fluid velocity while temperature, fluid concentration, and microorganism profiles show diminishing impacts. Applications of magnetic force offer a novel way to stabilize the flow, slow down the passage of fluid particles, and accelerate temperature, concentration, and microbe profile changes. Majeed et al. (Majeed *et al.*, 2024) incorporates the effect of chemically active stretchable surface on NHD bioconvection of nanoparticles. His findings suggest that while magnetic field slow down transport, buoyancy effects speeds it up. Thermophoretic effects causes strong boundary layers of concentration and temperature. Khan et al. (Khan *et al.*, 2024) extended the study of a stretching sheet in context of Prandtl hybrid nanofluid. Hussain et al. (Hussain *et al.*, 2024) employed the AI-based LightGBM algorithm to study the effects of a periodic magnetic field on the mixed bioconvection flow through a porous cavity. The results show that cultivable microbe levels increase with $Le$ and $Pe$. Additionally, variation of $Le$ is more significant on oxygen concentration and motile density than that of flow field and temperature. Muhammad et al. (Muhammad *et al.*, 2024) explored MHD thermos-bioconvection under chemical reaction on a stretching sheet considering radiation effects. High

Ri results in reduced accumulation of microbes at the edge while enhancing heat transfer. Higher $Ra_b$ results in low nanofluid velocity. Almutairi (Almutairi, 2025) further incorporated the effects variable thermal properties, Brownian diffusivity and motile density to analyse bioconvective dynamics of micropolar nanofluid coupled with radiative effects and chemical reactions. Electro-MHD thermo-bioconvection of tri-hybrid nanofluid is studied by (Jawad *et al.*, 2025). Amudhini et al. (Amudhini and De, 2025), however, observed the tetra-hybrid nanofluid is shown to transfer more heat and mass than the trihybrid one on investigating the MHD bioconvection of an unstable tetra-hybrid nanofluid via a permeable rotating disk wherein the authors considered various slip boundary conditions, Soret-Dufour effects, and variable viscosity and thermal conductivity. Ali et al. (Ali *et al.*, 2025) studied stagnation point flow in a bioconvection nanofluid on a rotating permeable disk. High inertia coefficient and porosity variable augments the axial velocity field but dampens the radial and azimuthal counterpart. The axial and azimuthal velocity profiles decline when the stretching parameter and velocity ratio are increased. Evidently, the motile microbes increase the *Pe* and *Le* for bioconvection. Reddy et al. (Reddy *et al.*, 2025) investigate the two-phase MHD flow of a nanofluid across a stretched sheet using the homotopy analysis method. The upward velocity profile is observed to increase when porosity parameter, magnetic parameter, and suction parameter is high, while the opposite trend occur as fluid deviates from Newtonian behaviour. Microbe and mass transmission is damped by the Brownian motion, and chemical reaction. Over a porous, sloped, and stretched cylindrical surface, Bafe et al. (Bafe, Firdi and Enyadene, 2025) investigate the MHD bioconvection of non-Newtonian nanofluids. The results show that suction/injection, inertia coefficient, porosity, and magnetic field intensity decreases the velocity while raising the system temperature. Surface heat transfer is improved by increasing the *Bi* and thermal relaxation time. Skin friction is decreased by non-Fickian mass flux and microbial Brownian motion while it increases for other parameters. There are some recent studies that investigated thermos-bioconvection under various combination of the above settings (Farooq and Imran, 2025; Mohamed *et al.*, 2025; Sohail *et al.*, 2025).

**Fig 5.** Schematics represents a physical model describing a near-wall transport of nanoparticles and microorganism in a bioconvective fuel cell configuration (adapted from (Nima *et al.*, 2020)).

| Name | Colour | Size | Shape | Flagellar details | Motility (µm/s) | Ref |
|------|--------|------|-------|-------------------|-----------------|-----|
| *Chromatium okenii (Bacterium)* | Purple Sulfur Bacterium | 3–10 µm | Rod-shaped cells | One or more polar flagella | 50-200 | (Di Nezio *et al.*, 2023) |

*Corresponding email: jdhar.me@nitdgp.ac.in

| *Protoceratium reticulatum* | Dark Brown | 35–65 µm | Oval | Two flagella | 100-400 | (Erga, Olseng and Aarø, 2015) |
|---|---|---|---|---|---|---|
| *Alexandrium minutum* | Green-Brown | 17–25 µm | Round to slightly oval | Two flagella | 100-500 | (Balech, 1995; Hallegraeff, 2003) |
| *Karenia brevis* | Light Brown | 20–50 µm | Flattened, oval to irregular | Two flagella | 100-500 | (Steidinger and Ingle, 1972; Kamykowski, Milligan and Reed, 1998; Magaña, Contreras and Villareal, 2003) |
| *Tetraselmis sp.* | Green | 5-25 µm | Oval or slightly elongated | Four equal length flagella | 100-400 | (Norris, Hori and Chihara, 1980; Arora *et al.*, 2013) |
| *Heterosigma akashiwo* | Yellow-Green to Golden-Brown | 15–30 µm | Oval to irregular | Two unequal flagella | 200-600 | (Bearon, Grünbaum and Cattolico, 2006) |
| *Chlamydomonas reinhardtii* | Green | 10 µm | Spherical | Two anterior flagella | 100-1000 | (Polin *et al.*, 2009; Avasthi *et al.*, 2014) |
| *Chattonella marina* | Yellow-Brown | 20–80 µm | Oval to elongated | Two flagella | 200-600 | (Tiffany *et al.*, 2001; Harvey, Menden-Deuer and Rynearson, 2015) |
| *Pyramimonas sp.* | Green | 5–20 µm | Pyramid-like or oval | Four- Eight flagella | 300-700 | (Bhuiyan *et al.*, 2015) |
| *Euglena gracilis* | Green | 20–100 µm | Elongated, spindle-shaped | Flagellar and Euglenoid movement | 200-1200 | (Lukáčová *et al.*, 2023) |
| *Dunaliella salina* | Green but Orange-Red under high salinity | 5–25 µm | Ovoid to elongated | Two flagella | 200-500 | (Craigie and McLachlan, 1964; Borowitzka, 1990) |
| *Karenia brevis* | Yellow-Brown | 20–40 µm | Flattened, oval to oblong | Two flagella | 100-500 | (Naar *et al.*, 2002; Magaña, Contreras and Villareal, 2003) |

**Table 3:** Details of some motile microorganisms commonly used in bioconvection experiments and their characteristics.

Although there are plethora of studies delving into the interaction of porous media and bioconvection under the impacts of various external factors and their combinations, critics to such studies have opined that such studies only have theoretical interests while relevance to practical situations may not be realizable. A major criticism of such studies is whether bioconvection can be sustained under strong porous confinement and whether the cells can survive various external factors, such as magnetic fields, steep temperature gradients, and significant constrictions within porous systems. Additionally, dispersion in porous media is generally known to suppress instabilities (Dhar *et al.*, 2022), raising an important question: can bioconvection persist in highly dispersive porous environments? Despite these concerns, several studies have even been conducted in non-porous configurations (open water systems) to investigate bioconvection flows, some of which we review below. Kotha et al. (Kotha *et al.*, 2020) studied the MHD flow and heat and mass transfer processes over a vertical plate subjected to heat production or absorption for a water-based nanofluid containing gyrotactic microorganisms. It was shown that the effects of magnetic field intensity on heat and mass transfer lead to decreased flux rates and velocity profiles of mobile microorganisms in the fluid medium. The temperature distribution, drag stress rate, and motile microbe profiles are all improved by the magnetic parameter. Kuman et al. (Kumar and Srikanth, 2021) use an analytical method to explore thermo-vibrational convection in a suspension of active (gyrotactic) swimmers. The main objective of this research is to use an over-stability investigation of thermo-vibrational convection within a well-defined mathematical and physical framework to characterize the heat transfer events in a gyrotactic suspension. A linear stability analysis provides a detailed description of the significant scientific and mathematical insights that this thermo-vibrational convection offers. The results show how temperature gradient and vertical vibration affect the gyrotactic patterns analytically. Temperature variation across the surfaces destabilizes both oscillatory and non-oscillatory types of convection in the system, reducing bioconvection. Cui et al. (Cui *et al.*, 2022) examines the effect of MHD bio-convective flow and stagnation point of bio-nanofluid containing motile microorganisms on a solid plate. The primary result reports an intensification of the velocity profile and velocity boundary layer thickness for higher temperatures, solute concentrations, and microbe *Gr*. The temperature profile becomes enhanced when the heat generation/absorption parameter values rise. The velocity and temperature field increase with increasing nanoparticle volume fraction values; however, the concentration distribution exhibits a declining trend. Microorganisms' density dampens for increasing values of *Pe* and *Sc*. Waqas et al. (Waqas *et al.*, 2022) present a model for the flow of radiative tangent hyperbolic nanofluid across a Riga plate in the presence of gyrotactic microorganisms. The results show that the momentum boundary layer grew with an increase in the modified Hartman number and mixed convection parameter; but dropped with an increase in the Weissenberg number and the second-order velocity slip parameter. A little increase in the thermophoresis parameter and Biot number values leads to an upward volumetric nanoparticle concentration trend. The thermo-bio-convection behaviour and entropy generation of gyrotactic microorganisms in a square enclosure with two square heat sources is studied by Izadpanah et al. (Izadpanah *et al.*, 2024). At high thermal Rayleigh numbers (Ra$_t$ = $10^5$), the average Nusselt number increases while average *Sh* decreases as the bio-convection Rayleigh number increases. Therefore, significant concerns remain regarding the feasibility of bioconvection in such diverse and constrained configurations. A promising alternative to Darcy-scale theoretical models is pore-scale simulations, which should not only capture the geometry of void spaces and their influence on cell swimming and collective behaviour but also account for the finite size of the cells and the increasingly dominant cell–cell interactions in small pore environments. Consequently, theoretical investigations must aim to delineate the parameter regimes and conditions under which bioconvection remains practically achievable.

*Corresponding email: jdhar.me@nitdgp.ac.in

*Corresponding email: jdhar.me@nitdgp.ac.in

***Theoretical background:*** Numerous theoretical and numerical studies have explored bioconvective phenomena under a variety of conditions, with the governing equations in each case tailored to the specific physical scenario. Here we present the most general form of the governing equations for bioconvection through a Darcy-scale porous medium for a Newtonian base fluid filled with nanoparticles and subjected to a magnetic field and heat source. Further variations such as non-Newtonian rheology or complex geometries can be readily extrapolated from this general equation framework. Readers interested in such specialized cases are encouraged to refer to the respective studies for detailed formulations. The governing equations describes the hydrodynamics in a porous medium with Darcy, Brinkman and Forchheimer model as

$$\frac{\rho_b}{\varepsilon^2}\frac{D\boldsymbol{u}}{Dt} = -\nabla p_e + \frac{\mu_b}{\varepsilon}\nabla^2\boldsymbol{u} - \frac{\mu_b}{K}\boldsymbol{u} - \frac{\rho_b}{\sqrt{K}}\frac{1.75}{\sqrt{150\varepsilon^3}}\boldsymbol{u}|\boldsymbol{u}| - \sigma_b B_0^2 u\hat{x} + \boldsymbol{g}[\gamma\Delta\rho m - \rho_b\beta_b(T - T_{ref})]$$

$$\nabla \cdot \boldsymbol{u} = 0$$

Here the last three terms describe the magnetic body force and buoyancy forces that depends on the microorganism's number density and temperature change. The above equations do not consider the effect of cell sedimentation and rotation or strain in the ambient flow. It is noteworthy that non-porous configurations have also been considered for such studies wherein the Darcy, and Forchheimer terms are excluded from the above governing equations. The temperature field is described by the energy balance equation as

$$\left(\rho c_p\right)_b \frac{DT}{Dt} = k_b \nabla^2 T + Q'''$$

Here $k_b$ is the thermal conductivity that is usually a function of the nanoparticle concentrations and porous medium characteristics. Through the term $Q'''$, the effects of internal heating, or incorporation of Brownian or thermophoretic effects can be incorporated. The microorganism dynamics is governed by the species balance equation

$$\frac{\partial m}{\partial t} + \frac{\partial}{\partial x}[(u + \tilde{u})m] + \frac{\partial}{\partial y}[(v + \tilde{v})m] = the them$$

Here $\tilde{u}$ and $\tilde{v}$ are the average cell swimming speed given by $\frac{bW_c}{\Delta C}\frac{\partial C}{\partial x}$ and $\frac{bW_c}{\Delta C}\frac{\partial C}{\partial y}$, respectively. For bacterial suspension the impact of oxygen concentration ($C$) is accounted for through the equation

$$\frac{DC}{Dt} = D_c \nabla^2 C - \mathcal{R},$$

where $\mathcal{R}$ is a term that implies the oxygen utilization rate. Here we note that some articles dealing with algae or phytoplankton instead of bacteria employs a separate balance equation for nanoparticle concentration ($\phi$) having a similar form as that of oxygen concentration balance where $\mathcal{R}$ implies the chemical reaction rate. Many other studies have included the effect of non-Newtonian fluid, radiation effects, chemical reaction, viscous dissipation that would mean incorporation of some additional terms to the above equation set. A concise review of various mathematical models for bioconvection in open waters and porous media can be found elsewhere (Barsanti *et al.*, 2025).

While the boundary conditions are case specific to various problems, usually no flux for heat, microorganisms and oxygen are applied across the walls, and magnetic field is usually applied perpendicular to the major flow direction. The meaning of the symbols used in the above formulation can be found in Table 1. Here we note that although the above formulation is numerically tractable and exhibits intriguing bioconvective patterns, we need to consider the impracticality of sustained bioconvection through porous medium in most of the above theoretical configurations presented above. The major reasons to doubt such an active phenomenon physically occurring owes to multiple reasons: the survivability of the cells in drastic temperature ranges, the effect of strong magnetic field on the cell population, the effect of porous media and its dispersion on the initiation and stability of the bioconvective plumes and the impact of low permeability of the medium for free swimming of the cells.

## 4. Blooms in porous medium

As previously noted, bioconvection and algal blooms are often sequential and synergistic phenomena. In this context, the study of bioconvection within porous media naturally extends to the examination of algal blooms in such environments. The tortuous structure of porous media offers a unique setting for exploring how these structures interact with bloom forming species, providing deeper insights into the mechanisms that govern both bioconvection and bloom dynamics in confined spaces. Intertidal rock pools, a form of porous environment along sea coastline (Bauer *et al.*, 2024), show distinct seasonal as well as diurnal variations in temperature, pH, $P_{CO_2}$ and $P_{O_2}$ signalling such that their porous entrapments are conducive habitats for bloom forming species (Morris and Taylor, 1983; Bauer *et al.*, 2024). Intertidal rocky shores exhibit algal blooms due to high nutrient accumulation despite strong grazing from other organism (Masterson *et al.*, 2008). The responses of two bloom-forming benthic dinoflagellates in intertidal systems, *Bysmatrum gregarium* and *Amphidinium carterae*, to changes in salinity and prolonged darkness are evident in the study by Rodrigues et al. (Rodrigues and Patil, 2022). The results indicate that low salinity caused by monsoonal rainfall may be able to control pre-monsoon benthic dinoflagellate blooms in intertidal systems along Indian coasts because it inhibits photosynthesis, hinders cell growth, and does not cause cyst formation as an adaptive survival strategy. Another coastline ecosystem is the sandy sediments which are known to home algal blooms (van Luijn *et al.*, 1999; Shen *et al.*, 2013). Submarine ground discharge into the sandy beaches at coastal regions in India is estimated to be around ~37m³/yr which contains nutrients, and trace metals. The porous ecosystem acts as a biogeochemical reactor, thereby increasing the chances of algal bloom formation (Bhagat and Kumar, 2022). In fact, harmful algal blooms (HABs), specifically those of the toxic diatom *Pseudo-nitzschia*, can originate offshore within the porous subsurface before appearing in surface coastal waters suggesting that subsurface populations can serve as a source for seeding surface HAB events (Seegers *et al.*, 2015; Piontkovski *et al.*, 2017). In an artificial environment, it is observed that *Microcystis aeruginosa* can break through a packed porous columns, unaffected by variations in ionic strength, media size, flow rate, or the presence of dissolved organic matter implying the transport and retention behaviour of the cells in a porous environment (Zhao *et al.*, 2019). In algae-based industries, porous medium plays a pivotal role to modify the photobioreactor (PBR) design to enhance biomass growth and simulate an artificial bloom formation (Podola, Li and Melkonian, 2017; Legrand, Artu and Pruvost, 2021; Carbone and Melkonian, 2023). Although not directly grown within the porous matrix, interactions of algal bloom and porous systems may occur through ground water discharge and membrane fouling scenarios. Algal bloom and porous subsurface interacts through the groundwater discharge that release essential nutrients into the shallow marine coastlands (sandy environments as stated previously) thereby promoting algal growth and sustenance

*Corresponding email: jdhar.me@nitdgp.ac.in

(Brookfield *et al.*, 2021; Su *et al.*, 2025). Biofouling due to algal organic matter release from algal blooms during ultrafiltration management has caused serious challenges in its technology (Mustafa *et al.*, 2022). Both ceramic and polymeric membranes, a porous substrate, exhibit reduced permeability due to pore-blocking and subsequent gel-layer formation making the fouling irreversible (Zhang and Fu, 2018; Malaguti *et al.*, 2022). Since these two aspects are not direct consequence of a porous environment harbouring an algal bloom, we have not detailed it in this review.

Research presently aims towards modified lime-ceramic sand-lake sediments to suppress the formation of algal blooms within its porous environment by exploiting the capabilities of flocculation, hydrodynamic conditions and local bridging effect (Xia *et al.*, 2022). Additionally, synthetic river sediments showed efficient removal of nitrogen, phosphorous and other metals (Liu *et al.*, 2019; Guo *et al.*, 2022). Lake bottom sediments was modified by Xia et al. with ceramsite sand and lime that acts as algal precipitation agent and control the bloom of *Microcystis aeruginosa* (Xia *et al.*, 2021). Besides sand and rock-wood habitats, biochar acts as another porous ecosystem which finds extensive applications in algal blooms specially to suppress the bloom formation due to its excellent capability of nutrients sorption from the surroundings (Cheng *et al.*, 2022; Kończak and Huber, 2022). Efficient nutrient removal from the surrounding thereby leads to reduced nutrient availability to the algae in open waters (Cheng *et al.*, 2023). A review on biochar capability to hold nutrients may be found elsewhere (Vasseghian, Nadagouda and Aminabhavi, 2024). Biochar, therefore, limits the bloom growth within its porous environment and leads in mitigating eutrophication (Huang *et al.*, 2023; Liu *et al.*, 2023). In addition, electrostatic interaction among the biochar particles and algal cells also enhances the formation of larger aggregates that settles rapidly (Gorovtsov *et al.*, 2020; Lin *et al.*, 2023). Porous granules of dolomite are further reported to suppress algal bloom formation, transforming the dark green colour of the water containing raw algae to transparent colourless (Huh *et al.*, 2017). Although the occurrence, suppression or prevention of harmful algal blooms (HABs) in porous environments is less well-documented compared to surface waters, the potential exists for porous medium to form or hinder microbial blooms under the right conditions. This is an area of increasing interest, especially regarding bioconvection and the unique flow dynamics created within porous materials that might influence algal growth and its ecological impacts. Despite the above studies underscoring the importance of interactions between algal blooms, bioconvection, and porous environments, there remains a lack of dedicated research analysing how specific characteristics of porous systems influence these dynamics. Such investigations would hold substantial relevance at industrial scales, particularly in applications like carbon sequestration and biofuel production, where controlled algal growth and transport through porous media could offer significant advantages.

*Corresponding email: jdhar.me@nitdgp.ac.in

## 4 Genetic engineering towards modulating bioconvection

In a recent study, cell motility and induced bioconvection has been exploited to design a stirless photobioreactor that exhibit energy efficient operation (Martinez Carvajal, Taidi and Jarrahi, 2024). The ability of *Chlamydomonas reinhardtii* to swim and enhance the biomixing in the photobioreactor system signals an interesting shift in the biomass industry to use highly motile algal strains as mixing agents replacing mechanical mixing. In addition, porous photobioreactors have recently found tremendous potential to enhance biomass production (Podola, Li and Melkonian, 2017). Thus, highly motile algal strains can be used as a model species for investigating bioconvection in porous medium which may find novel industrial applications in the future. Creating highly motile algal strains demand genetic engineering. Genetic alterations of algal species have been used to modify various attributes and phenotypes for desired requirements. One of the key aspects of biofuel industry is to enhance biomass growth and lipid content. Genetic engineering has been used to increase the growth rate of algal species used for biomass, bioethanol, biohydrogen and biodiesel production (Schenk *et al.*, 2008; Narravula Raga *et al.*, 2021). Routine genetic transformation is carried out on few model organisms, one of which is *Chlamydomonas reinhardtii* (Tran and Kaldenhoff, 2020). *C. reinhardtii* can exist in both haploid and diploid states, enabling straightforward genetic manipulations. Researchers can perform tetrad analysis, facilitating the study of genetic segregation and linkage. Genetic analyses have identified several genes involved in flagellar length regulation and beating patterns. Some recent studies have simultaneously conducted genetic alteration to control flagellar motion and, in turn, cell motility, thus, impacting the population dynamics of the species (Qiao *et al.*, 2022; Sun *et al.*, 2022; Tran *et al.*, 2022). Intriguingly, *Chlamydomonas reinhardtii* also exhibit bioconvection (Sato, Sato and Toyoshima, 2018; Martinez Carvajal, Taidi and Jarrahi, 2024), thus, acting as a potential subject to genetically engineer a highly motile version of itself (Huang, 1986). Schierenbeck et al. generated high light-tolerant mutants of the microalga *C. reinhardtii* through low-level mutagenesis and stringent selection wherein both mutants exhibited increased non-photochemical quenching (NPQ) rates and improved resistance to chemically induced reactive oxygen species (ROS) (Schierenbeck *et al.*, 2015). Baier et al. demonstrated that introns can significantly increase transcript abundance in *C. reinhardtii* through a process known as Intron-Mediated Enhancement (IME), a phenomenon observed in various eukaryotes that enhances gene expression post-transcriptionally (Baier *et al.*, 2020). Additionally, genetic engineering is routinely employed to augment economic output for microalgal biproducts (Sproles *et al.*, 2021; Gupta *et al.*, 2024). These studies make it evident that genetic modification is a viable approach to enhance and regulate the motility and bioconvective behaviour of microbial species. However, we emphasize that this review does not focus on genetic modification itself but rather aims to highlight its potential relevance. In the context of bioconvection-related applications, exploring genetic modifications could serve as a promising direction for future research.

*Corresponding email: jdhar.me@nitdgp.ac.in

## 5 Conclusion and Future Perspectives

The study of bioconvection and microbial bloom formation within porous media has advanced significantly, revealing complex interactions between microbial motility, fluid dynamics, and environmental heterogeneity. Recent research has expanded our understanding of how motile microorganisms, such as gyrotactic microbes, induce bioconvective patterns in nanofluid flows through porous structures. These studies underscore the importance of considering both microbial behaviour and the physical properties of the medium to fully comprehend the dynamics of bioconvection in such environments. In parallel, genetic engineering has emerged as a powerful tool to modify algal behaviour, particularly concerning motility and growth characteristics. Advancements in gene editing and metabolic engineering have enabled the enhancement of microalgae biomass quality, facilitating the production of high-value metabolites and biofuels. These developments not only improve the economic viability of microalgal applications but also contribute to environmental sustainability by optimizing carbon sequestration processes.

Despite these advancements, several challenges and opportunities remain:

1. *Mechanistic Understanding*: Further research is needed to elucidate the precise mechanisms governing or suppressing bioconvection in porous media, particularly under varying environmental conditions. This includes understanding the role of microbial interactions and the impact of porous structures on bioconvection sustenance and flow patterns.

*Corresponding email: jdhar.me@nitdgp.ac.in

2. *Modelling and Simulation*: Developing robust computational models that accurately represent the complex dynamics of bioconvection in porous media is crucial. Such models can aid in predicting system behaviour at pore-scale resolution and guiding experimental designs.

3. *Genetic Engineering Applications*: Expanding the molecular tools available for genetic engineering of microalgae will lead to enhanced cell motility, population emergence, product yields and accelerate the development of new applications. This includes the creation of microalgal strains with tailored motility and metabolic profiles suited for specific industrial processes.

4. *Environmental and Industrial Integration*: Translating laboratory findings to real-world applications requires addressing scalability issues and ensuring that engineered microorganisms perform effectively in natural or industrial settings. This involves optimizing growth conditions and assessing ecological impacts.

By addressing these areas, future research can further harness the potential of microbial systems in porous media, leading to innovations in biotechnology, environmental management, and sustainable energy production.

**Author Contribution:** SB compiled the literature review for bioconvection in porous medium and prepared Table 3, SM compiled the bioconvection in open waters and prepared Table 1 and 2, and JD conceived the idea, compiled the review on bloom in porous medium, the theoretical framework and genetic engineering of microbes. SB, SM, and JD co-wrote the manuscript.

**Research funding:** This work was supported by the Department of Science and Technology, Government of India [grant number SRG/2023/001253].

*Corresponding email: jdhar.me@nitdgp.ac.in

*Corresponding email: jdhar.me@nitdgp.ac.in

*Corresponding email: jdhar.me@nitdgp.ac.in

*Corresponding email: jdhar.me@nitdgp.ac.in

*Corresponding email: jdhar.me@nitdgp.ac.in

8.

*Corresponding email: jdhar.me@nitdgp.ac.in